\documentclass{elsart}

\usepackage{psfig}
\usepackage{epsfig}
\usepackage{graphicx}

\newcommand{\AmS}{{\protect\the\textfont2
  A\kern-.1667em\lower.5ex\hbox{M}\kern-.125emS}}

\begin{document}
\begin{frontmatter}

\title{Prototype Tests for the CELESTE Solar Array Gamma--Ray
Telescope}

\author[Bordeaux]{
B. Giebels\thanksref{IKKE}},
\author[Toulouse]{R. Bazer-Bachi},
\author[LAL]{H. Bergeret},
\author[LAL]{A. Cordier},
\author[Perpi]{G. Debiais},
\author[X]{M. De Naurois},
\author[Toulouse]{J.P. Dezalay},
\author[Bordeaux]{D. Dumora},
\author[LAL]{P. Eschstruth},
\author[Lpc]{P. Espigat},
\author[Perpi]{B. Fabre},
\author[X]{P. Fleury},
\author[Lpc]{C. Ghesqui\`ere}
\author[LAL]{N. Herault},
\author[Toulouse]{I. Malet},
\author[LAL]{B. Merkel},
\author[Perpi]{C. Meynadier},
\author[AKA1]{M. Palatka},
\author[X]{E. Par\a'e},
\author[Bordeaux]{J. Procureur},
\author[Lpc]{M. Punch},
\author[Bordeaux]{J. Qu\a'ebert},
\author[Bordeaux]{K. Ragan\thanksref{ken}},
\author[AKA2]{L. Rob},
\author[AKA1]{P. Schovanek},
\author[Bordeaux]{D.A. Smith},
\author[X]{ J. Vrana\thanksref{Deceased}}

\address[Bordeaux]{CEN de Bordeaux-Gradignan, Le Haut Vigneau,
F-33175}       
\address[Toulouse]{CESR,  Toulouse, F-31029}                          
\address[X]{LPNHE, Ecole Polytechnique, Palaiseau, F-91128}  
\address[LAL]{LAL, Universit\a'e Paris Sud and IN2P3/CNRS, Orsay, F-91405} 
\address[Lpc]{LPC, Coll\a`ege de France,  Paris, F-75231}   
\address[Perpi]{GPF, Universit\a'e de Perpignan, F-66860} 
\address[AKA1]{Joint Laboratory of Optics of PU and Inst. of Physics
AS CR.} 
\address[AKA2]{Nuclear Centre, Charles University, Prague, CR.}
\thanks[IKKE]{e-mail: {\em giebels@cenbg.in2p3.fr}}
\thanks[ken]{Visiting from McGill University, Montreal, Canada.}
\thanks[Deceased]{Deceased.}

\begin{abstract}
The {\small CELESTE} experiment will be an Atmospheric Cherenkov detector
designed to bridge the gap in energy sensitivity between current satellite
and ground-based gamma-ray telescopes, 20 to 300 GeV. We present test results
made at the former solar power plant, Themis, in the French Pyrenees. 
The tests confirm the viability of using a central tower
heliostat array for Cherenkov wavefront sampling.
\end{abstract}

\end{frontmatter}

\section{INTRODUCTION}
\begin{figure}
\includegraphics[scale=0.9]{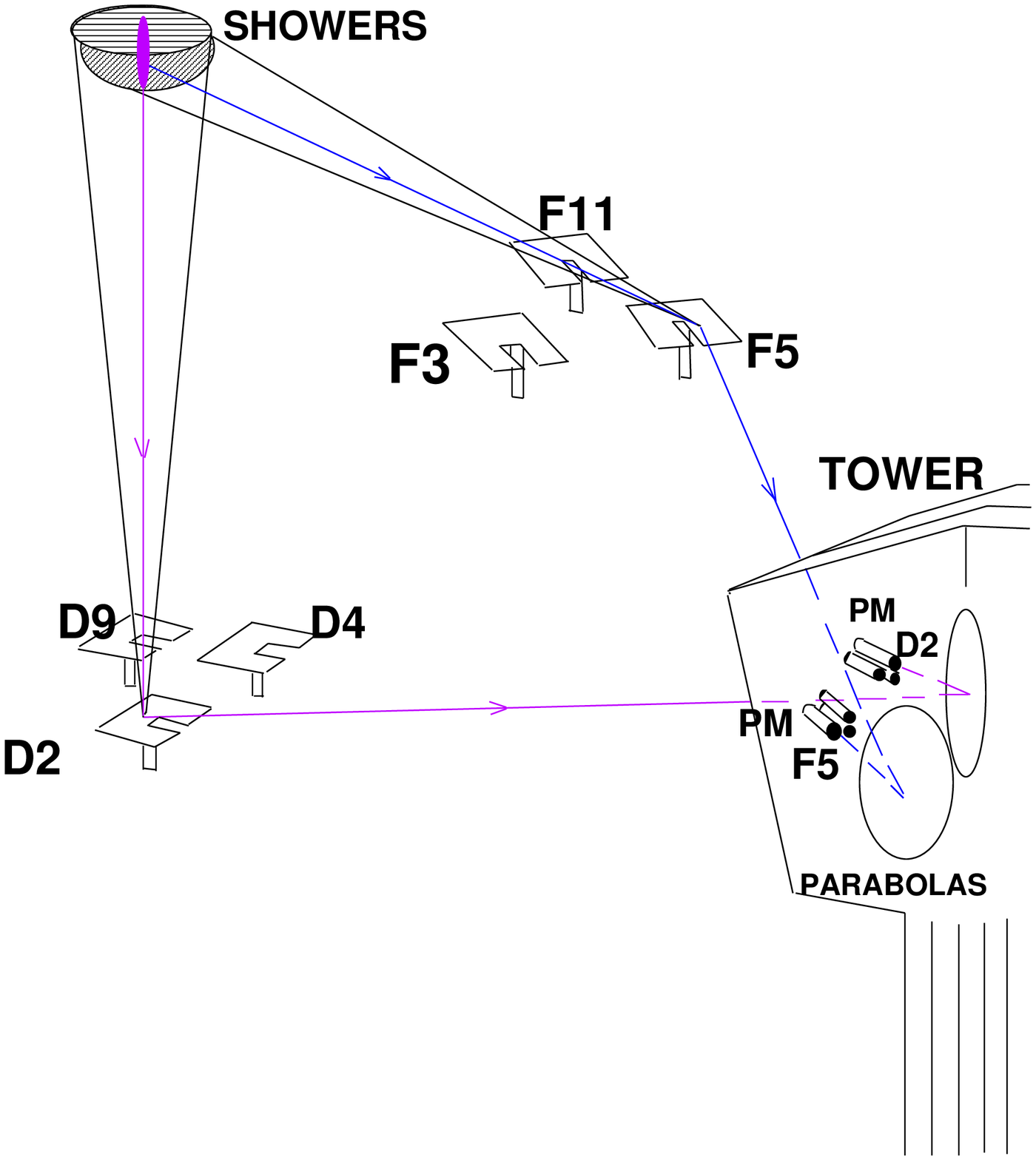}
\caption{  Principle of a ``solar farm'' Cherenkov gamma-ray telescope,
showing the six heliostats and the two searchlight mirrors used 
during tests at Themis in the Winter of 1996/1997. Heliostats and secondary mirrors are not to scale. The tower is 100 meters tall.}
\label{principle}
\end{figure}

The {\small CELESTE} experiment uses the
Electricit\a'e de France central receiver solar power plant at Themis 
(N. 42.50$^\circ$, E. 1.97$^\circ$, 1650 m. a.s.l.) as a gamma-ray telescope
\cite{proposal}.
The project is fully funded and will begin observations in early 1998
with 18 heliostats, growing to 40 in the following year.
Figure \ref{principle} sketches the principal of the {\small CELESTE} approach. This paper describes test results accumulated
during the design and construction phase. Emphasis is placed on Cherenkov
measurements made using six heliostats between October 1996 and February 1997.

\subsection{Science}
Active Galactic Nuclei (AGNs), pulsars, 
and supernova remnants 
are complex ``cosmic accelerators''.
High energy radiation dominates the power output
of certain classes of these objects \cite{vonM,Egpuls}.
Study of their high energy spectra provides insight into the origin of cosmic rays,
especially at the highest energies, and the nature of AGNs.

Photons from distant galaxies also probe the extragalactic medium:
gamma-rays incident on near infrared photons are above threshold for
$e^+e^-$ pair production and are thus absorbed, with approximately
one attenuation length for GeV to TeV photons traversing cosmological 
distances. 
In this way gamma-ray spectra provide information on galaxy formation and, indirectly, on the nature of dark matter \cite{ir}.

Around 1990 two breakthroughs revolutionized the field.
First, ground-based atmospheric Cherenkov detectors became sensitive, reliable
instruments above a few hundred GeV, led by the Whipple imager \cite{whipple}
and followed
by the Themistocle and {\small ASGAT} wavefront samplers at Themis
\cite{them,asgat}.  
Recent measurements by {\small CAT} and other imagers of flaring in Mrk 501 underscore
the rapid improvement of ground-based detectors \cite{cat}.
Second, the {\small EGRET} instrument on the
Compton satellite measured the spectra of over 150 point sources
and mapped the galactic diffuse gamma-ray emission, in the
energy range $0.1 < E_\gamma < 10$ GeV \cite{Egcat}.
The energy range
currently inaccessible either by satellite or ground-based
detectors, $20 < E_{\gamma} < 200$ GeV, is particularly rich with information on the acceleration and absorption
processes. Bridging this energy
gap is a key scientific goal.

\subsection{Solar Arrays as Atmospheric Cherenkov Detectors}
The Atmospheric Cherenkov Technique has been described extensively elsewhere:
see for example \cite{weekes,whipple,them,asgat,cat}. 
A few points bear repeating.

The minimum energy threshold of a Cherenkov telescope is limited by
accidental trigger coincidences induced by photons from the night sky. 
For a constant diffuse night sky light $\phi$
(photons per unit time, area, and solid angle),
and a coincidence time gate $\tau$ the threshold scales as
\begin{equation}
\label{thresh_form}
E_{threshold} \propto \sqrt{ {\Omega \tau \phi} \over {A \epsilon}}
\end{equation}
where
$\Omega$ is the solid angle seen by a phototube and $\epsilon$ is the photon collection efficiency. 
Current telescopes have pushed $\tau$ and $\Omega$
to their practical limits. Until technological progress
improves quantum efficiencies and hence $\epsilon$, lower
energy thresholds require larger mirror areas, $A$.

Solar heliostat arrays offer the advantage of a large mirror area available
without additional construction costs. 
The optical configuration is similar to that of 
multiple mirror wavefront samplers such as Themistocle. 
The main difference is the common focus of all heliostats is situated at
        the top of the tower. Secondary optics disentangle the light coming from  each
        of the heliostats, as sketched in figure \ref{principle}.
The {\small STACEE} collaboration is building a similar detector in 
the United States \cite{stacee}.

Imagers are superior to samplers above 200 GeV, since the fine sampling of
their cameras allows excellent hadron rejection. 
But below 100 GeV,
hadron backgrounds are naturally suppressed because the Cherenkov yield
of the hadron showers decreases faster than that of the gamma showers.
Hence, the advantage of the imagers in this regard is less.

At low energies, the background due to cosmic ray electrons is 
greater than above 200 GeV due to an energy spectrum
steeper than for hadrons. Since electron
induced showers are practically indistinguishable from gamma-ray
showers, angular resolution is the key to suppressing the electron background.
But angular resolution, especially at lower energies, tends to be
limited by shower development rather than by instrument response.
Scattering in the first few generations of the shower and deflection
of low energy secondaries in the geomagnetic field are the factors limiting
angular precision. Hence, a high-granularity imager does not necessarily
outperform a wavefront sampler. Both should achieve angular resolutions around
        $0.1^\circ$.

Sections 2 and 3 describe the apparatus and the main results of the
        prototype tests. Based on these results we   compare  sampling  arrays  and
        imagers for the low energy domain in section 4.

\section{EXPERIMENTAL APPARATUS}
We chose two well-separated (156 m) groups of three nearby (20 m) heliostats
each to test the strategy adopted for {\small CELESTE}: a long lever-arm
for event reconstruction combined with short path-length differences
within subgroups to simplify the trigger electronics (figure \ref{principle}).

\subsection{Photomultiplier tubes}
The Philips XP2020 used for these tests
 is a fast 12-stage 2-inch photomultiplier that has been
used extensively at the Themis site. The photocathode is
bialkili and overall it is very similar to the 8-stage
XP2282B we have selected for the 40-heliostat experiment.
The bases were of the ``C'' type, supplied by Philips.
We measured their gain via the single photoelectron peak in
both charge and voltage (see \cite{rasmik} for a discussion). 
All gains were measured with 15 meter RG-58 cables as were used
for the experiment. The average gain was $10^6$ and
anode currents due to night sky light were typically 70 $\mu$A. 
The phototube signals were amplified by 10 using a LeCroy 612AM NIM
module.  
 
\subsection{Readout}
Electronics and computers were installed in a counting house just
beneath the secondary optics.
For the tests we used three Struck DL515 Flash ADC's. 
The DL515 has four channels that each sample at 250 MHz.
Interleaving a signal over four channels provides 1 ns sampling.
These
VME modules were controlled by a Motorola 68040 processor running
under the Lynx operating system. We recorded
400 one-nanosecond 8-bit voltage samples from each Flash ADC for each
trigger. Trigger rates were under 10 Hz and deadtime due to
readout was less than 5\%.

A typical Cherenkov event is shown in figure \ref{typ_evt}.
As only three Flash ADC's were available at that time we   
 recorded the six phototube signals by AC-coupling the analog 
sum of three phototubes to a single Flash ADC, as sketched
in figure \ref{mixfadc}. This increased the night sky light in the data. Short delays
(8 and 16 ns) were added to separate Cherenkov
pulses from each another. 
The average overall conversion factor was $1.8 \gamma e$ per digital
count, with a $\pm 15\%$ channel-to-channel variation. (The full-scale
detector will have a conversion factor of $0.4 \gamma e$ per count
to enhance sensitivity to small pulse heights.)  
The electronic pedestal was negligible. With AC-coupling, the pedestal of the night sky light fluctuations  was
$1.2 \pm 1.0 \gamma e$ per 10 ns. 
We studied the precision of the time and charge measurements made
with the FADC's by superimposing simulated phototube pulses onto
real night sky data and then reconstructing the pulses.  
For an amplitude of 3 photoelectrons we found 
$\delta(\Delta t) \simeq 0.5$ ns and $\delta V/V \simeq 15$\%.
The precision improves with increasing pulseheight.

In addition to the VME system,  
we also used Camac ADC's (LeCroy 2249SG), 
TDC's (LeCroy 2228A), discriminators (Phillips 7106), and scalers (LeCroy 2551)
with PC/GPIB readout for each phototube. 
These provided cross-checks of the Flash ADC results.

\subsection{Trigger}

{\em Digital trigger:}

For each phototube an amplifier output 
was DC-coupled to a discriminator circuit with a 20 ns output gate.
A multiplicity requirement ({\em e.g.}, five of the six channels
above threshold) stopped the Flash ADC's and prompted the
acquisition computer to read the Flash ADC memories.

Threshold settings were the same for all six phototubes.
The DC-coupling causes a baseline voltage $V_{sky}(b)=bgeR$ that varies
 with the night sky light induced rate of $b$ photoelectrons
per nanosecond
at the photocathode, where $g$ is the combined phototube and
amplifier gain, $R=50\Omega$ is the termination resistance,
and $e$ is the electron charge. 
We made frequent measurements
of the unamplified anode current which we used when
converting the trigger thresholds to photoelectrons.
A typical value of $V_{sky}$ was -30 mV.
The discriminator thresholds were -80 mV, 
which corresponds to 5 photoelectrons above the mean
night sky light level (except as noted).

{\em Hybrid Analog-Digital trigger:}

 Atmospheric Cherenkov telescopes operate at trigger thresholds limited by accidental triggers due to night sky light. For the {\small CELESTE} experiment the rate of single photoelectrons on an individual phototube
is high (600 MHz), and a goal of the full 40-heliostat experiment is to lower the effective trigger threshold to 3 photoelectrons on
each phototube, corresponding to gamma-rays of about 20 GeV.

\begin{figure}
\includegraphics[scale=0.9]{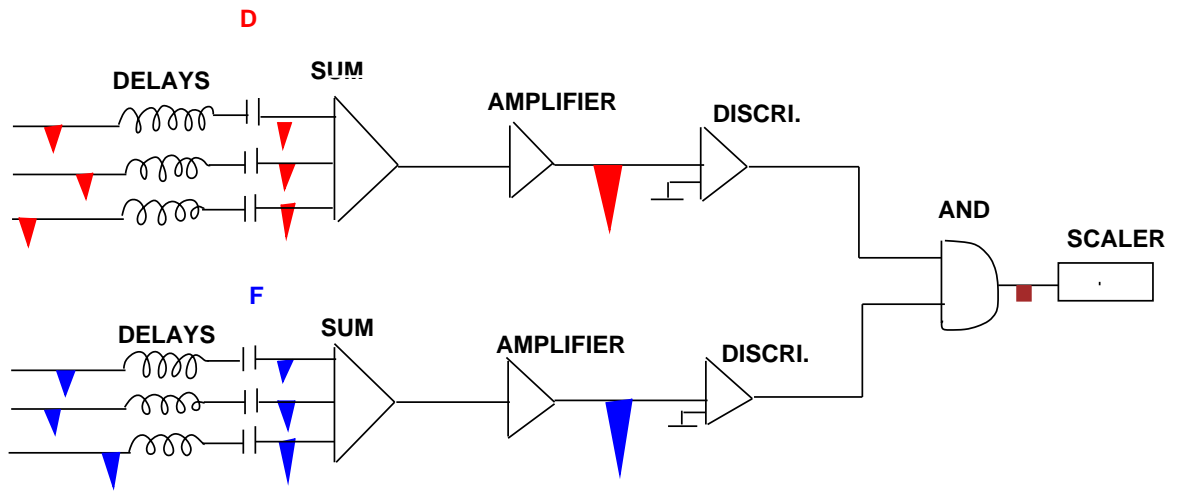}
\caption{  Hybrid analog-digital trigger layout for the six-heliostat
array.}
\label{trigsum}
\end{figure}

\begin{figure}
\includegraphics[scale=0.9]{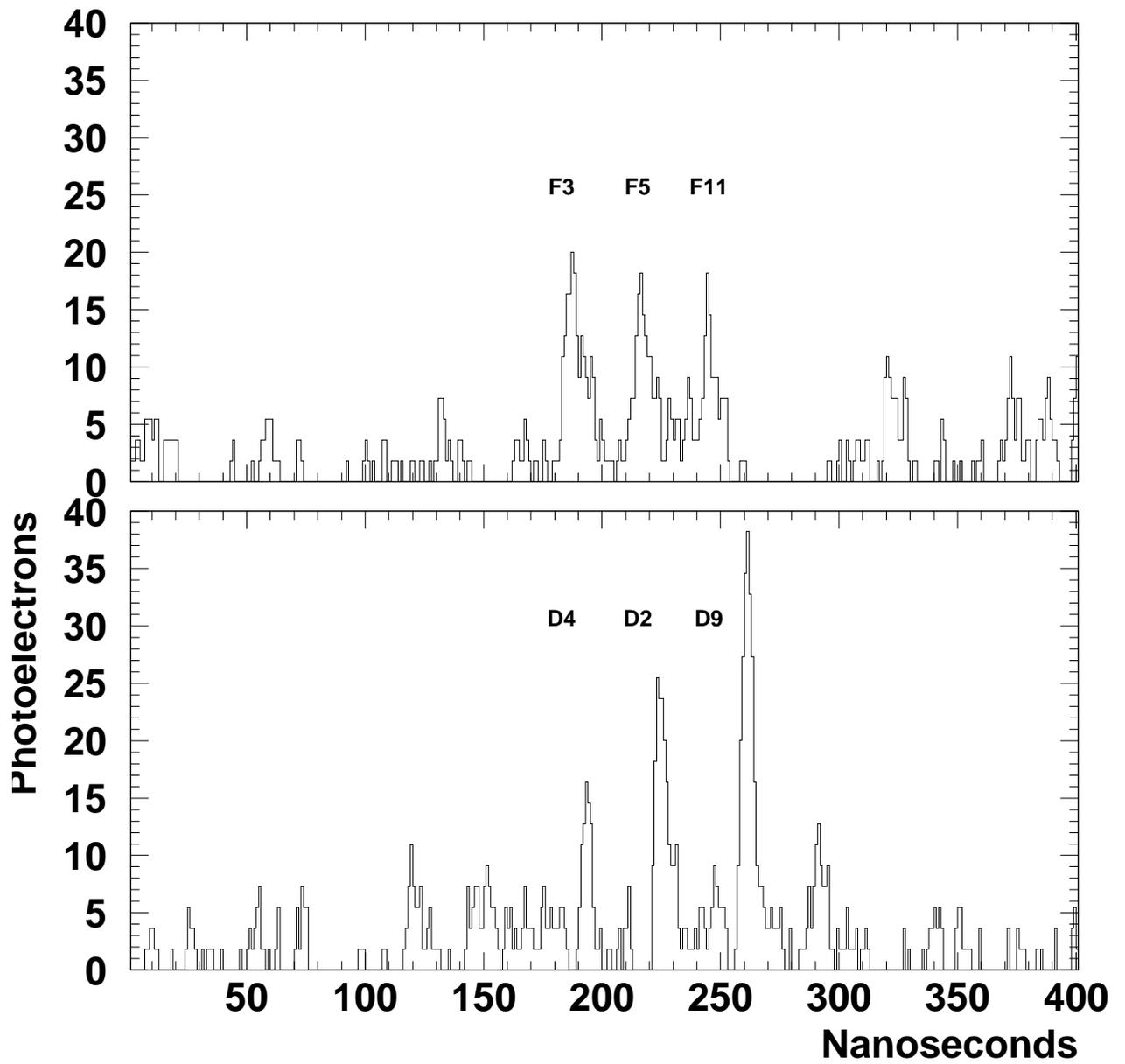}
\caption{  Typical Cherenkov flash as seen by the six phototubes
and recorded by two Flash ADC circuits. 
The labels are the heliostat names shown in figure \ref{principle}.}
\label{typ_evt}
\end{figure}

\begin{figure}
\epsfig{figure=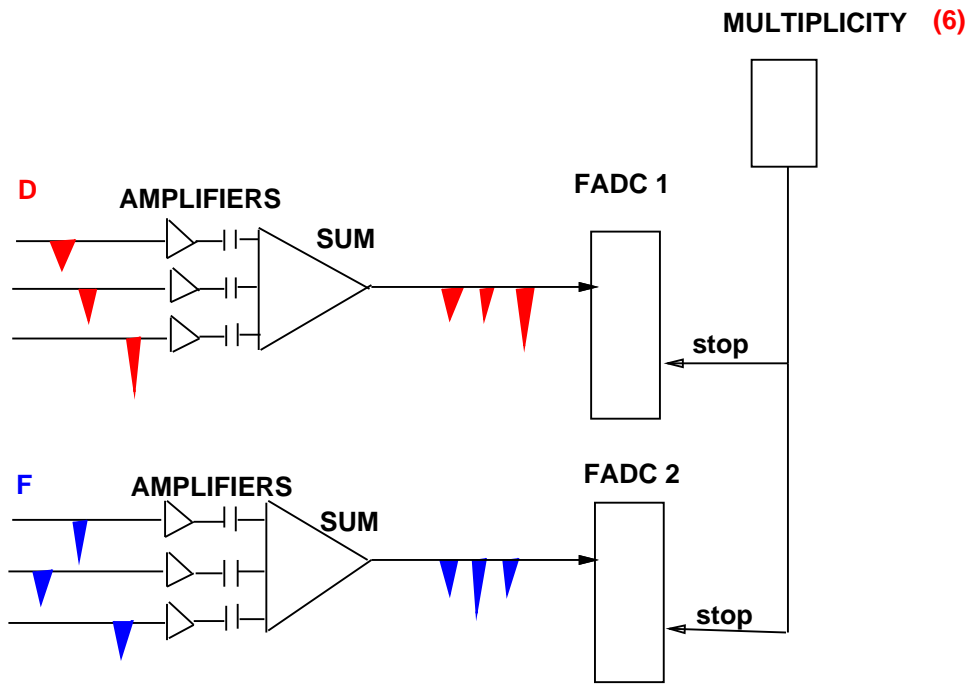}
\caption{  Scheme whereby six photomultiplier signals are
digitized using two Flash ADC channels.}
\label{mixfadc}
\end{figure}

The analog trigger sums the signals from the phototubes within  a  group  of
        heliostats,  and  the  sum  is  discriminated, as sketched in figure \ref{trigsum}.  The  fast   time
        structure of the phototube pulse and the synchronous secondary optics provide the smallest  possible
        $\tau$  (eq.  \ref{thresh_form}),  and  thereby  a  low   energy
        threshold. A purely analog scheme generates triggers which arise
        from, for example, phototube  afterpulses,  charged  particles  or  muons.  A
        coincidence of the discriminated outputs from groups  eliminates
        noise triggers which would occur using only  analog  sums.  This
        coincidence between analog sums is the hybrid analog-digital trigger.
 The principal reason that {\small CELESTE}
will use a hybrid analog-digital trigger is that it lowers the trigger threshold at which the rate due to Cherenkov flashes dominates the rate due to accidental coincidences of night sky light induced pulses. This allows operation nearer the effective 3 photoelectron per heliostat threshold, which should achieve a 20 GeV gamma-ray threshold.

Two three-channel analog summing circuits,  illustrated in figure \ref{trigsum}, were available during the February '97
tests, which we used for the measurements described in section
\ref{trg_sect}.
The key component is a passive inductive summer/splitter manufactured
for satellite radio-frequency repeaters \cite{mini}.
The bandpass is 500 MHz and the insertion loss is 5.6 dB.

\subsection{Heliostats}
\begin{figure}
\includegraphics[scale=0.65,angle=-90]{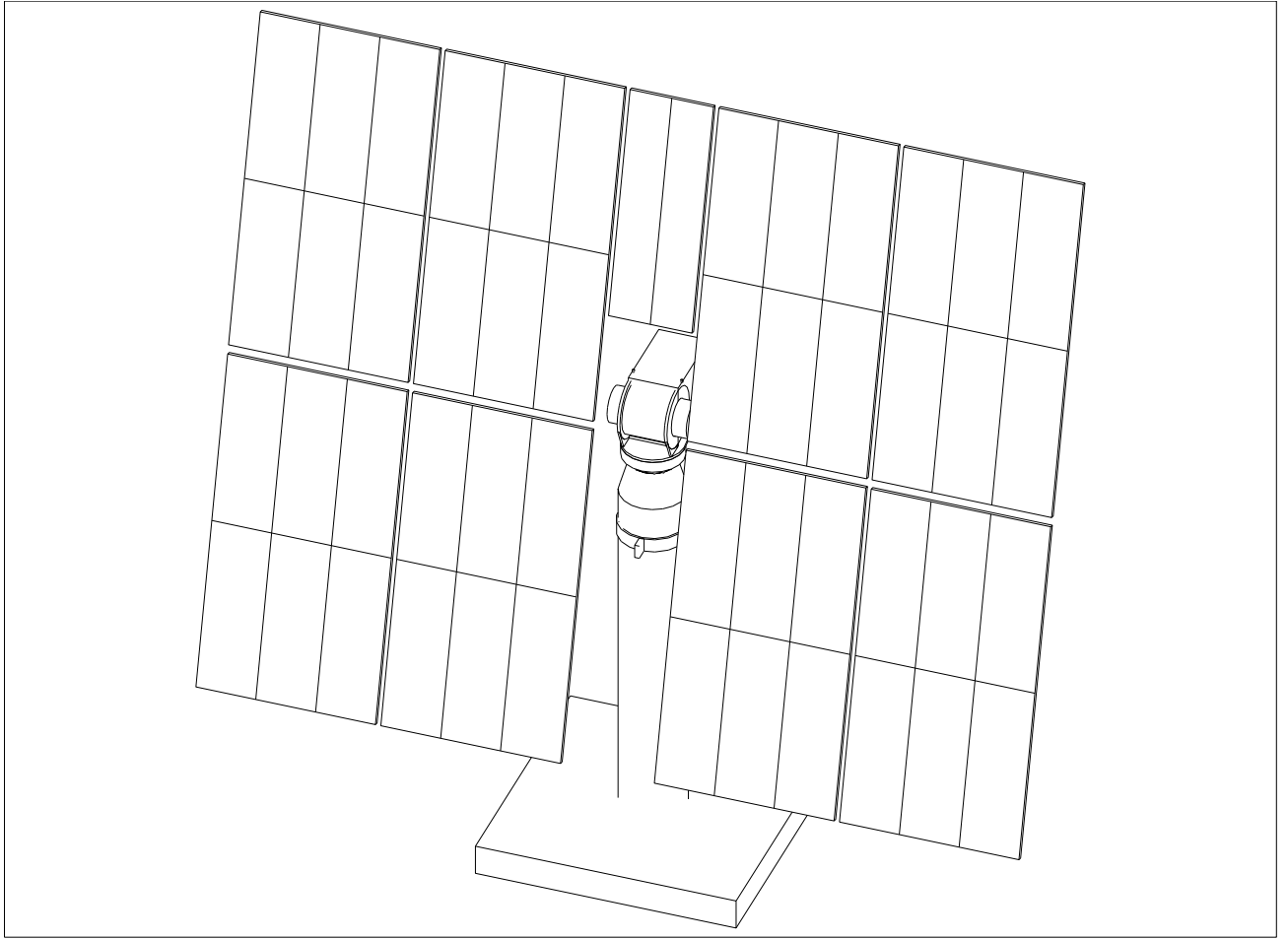}
\caption{  A single Th\a'emis heliostat mirror (7.3 by 8.8 meters)
on its alt-azimuth mounting.}
\label{hstat}
\end{figure}

Figure \ref{hstat} shows a single heliostat. 
The microprocessor that
controls the two DC motors of the alt-azimuth mount receives pointing
instructions from a central computer via a serial bus. The motor
step size is 0.14 mr. Heliostat alignments were adjusted by maximizing the
phototube anode current during star tracking, and were cross-checked
using stars at widely spaced points in the sky.
The spherical mirrors are silver back-coated with a surface area of
54 m$^2$.

\subsection{Secondary optics}
\label{nsl}
We aluminized two searchlight mirrors and mounted them in the
tower in the space formerly occupied by the solar heat receiver.
Each pointed towards a group of
heliostats. The mirrors are f/0.4 parabolas with an 150 cm
aperture and 90\% reflectivity. The field-of-view (``{\it fov}'') 
is thus in principle given by  the  angle  subtended  by  the  secondary
        mirror when viewed from a heliostat. The 
        heliostats we used are all approximately at 140 meters from the tower  so
        our {\it fov} was 1.5/140 = 11 mr (full width).
However, the short focal length gives rise to large off-axis
aberrations so that light falling near the mirror edge is lost.

We calculate the light collection efficiency by ray-tracing
techniques, up to the phototubes. We modeled the detailed heliostat structure,
and used the phototube positions in the parabola's
focal plane that were used during the experiment. 
Various measurements confirm the validity of
our calculations.

\begin{figure}
\includegraphics[scale=0.9]{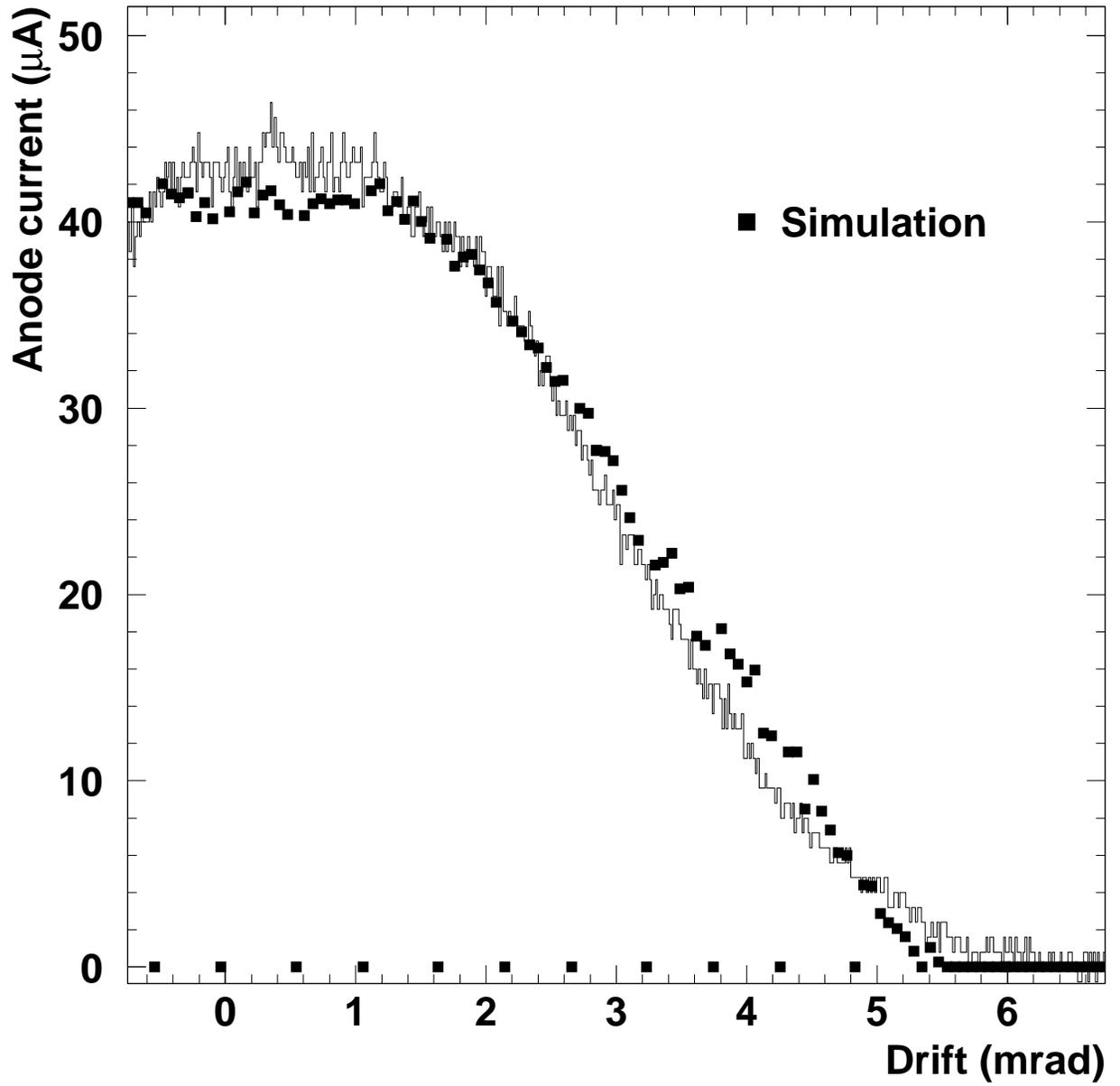}
\caption{  Drift scan of the star $\alpha$Leo. 
Data: measured phototube current. Simulation: ray tracing calculation
of the heliostat-parabola combination.}
\label{drift}
\end{figure}

{\em Star drift scans:}
We made two dimensional maps of the heliostat optical acceptance
by letting stars drift through the field of view and recording
a photomultiplier's anode current. (Phototube
high voltage was reduced to avoid excessive anode current.)
Figure \ref{drift} shows one such scan.
Superimposed is the ray tracing result, normalized to the
peak anode current. We see that the effective
{\it fov} is 7.5 mr (FWHM) for a point source. 

{\em Night sky light: }
We measured the night sky light $\phi$ (see equation \ref{thresh_form})
with an XP2020 phototube in a collimating cylinder whose field-of-view  was defined by diaphragms. 
This was done in order to make a {\it direct} measurement of the night sky
with a well-defined  {\it fov} and with a flat acceptance over the entire 
{\it fov}. This also avoided other noise such as albedo. 
The night sky intensity at
        zenith was found to be $$\phi = (300 \pm 45) \gamma_e /  m^{2} 
        \textrm{ }sr\textrm{ } ns. $$  If  we  assume  a  flat  
        spectrum from $310$ to $650 nm$  we can deconvolute the
intensity from the phototube quantum  efficiency  and  obtain  a value of 
$$\phi = (1.8 \pm 0.3)\cdot10^{12} photons /  m^{2}\textrm{ }  sr\textrm{ }s $$
 which is comparable to measurements at other dark sites\cite{stacee,rasmik}.
We also measured the light from the same  part  of  the  sky
        after reflection by a piece of heliostat mirror.  
The intensity decreased  by 20\%,  in  good  agreement  with 
the measured heliostat wavelength dependent reflectivity.   

Using ray tracing programs to calculate the overall
transmission of our optics, the above measurement leads us
to expect a photocathode
illumination of $b=0.43 \gamma e$/ns for no albedo. If we assume
albedo due to 20\% ground reflectivity, $b$ increases to $0.57 \gamma e$/ns.
The measured values were typically $b=0.5$, with $\pm 5\%$ variation
from night to night. After a heavy snowfall in January, the measured
illumination doubled. We reproduce this effect by assuming 90\% ground
reflectivity. The definitive secondary optics now being installed 
is less sensitive to ground albedo.

{\em Laser data:} The Themistocle \cite{them}  
laser illuminates the heliostat field with 4 ns pulses (rms) at $\lambda=330$ nm
via a diffuser placed five meters below the parabolic secondary
that observes the Eastern group of three heliostats.
We recorded 10,000 laser pulses with the Flash ADC's and the six
phototubes, and we compared the laser intensity with the results
obtained from Themistocle, corrected for the different detector
geometries. We obtain the same average laser intensity as they,
thereby validating our optics calculations for a source in the 
mirror focal plane, a test complementary
to the study of a star (point source at infinity) or the night
sky (diffuse source).

\subsection{Monte Carlo shower and detector simulations}
The measurements described in the next section are compared to
the results of a Monte Carlo simulation.
We generated $650,000$ proton air showers with ISU \cite{isu},
with
energy $E_p > 150$ GeV according to a power law energy spectrum
of the form
\begin{equation}
\label{proton_eq}
\phi_P(E) = 2.43({ E \over GeV})^{-2.78}/ (cm^{2} \textrm{ }sr\textrm{ } \textrm{ } GeV)
\end{equation}
over a disk of radius 300 meters centered at D2 and uniformly
within $2^0$ of zenith \cite{masakatsu}. In addition we simulated
$60,000$ helium showers using the spectrum from the same reference,
and $100,000$ lithium showers according to the measurements of 
\cite{barbara}. (As can be seen in figure \ref{altitude} the
contribution from lithium is small and so we have neglected the heavier
nuclei.)
Cherenkov photons reaching the heliostats are then propagated
through a detector simulation that includes the optical and 
phototube properties, allowing us to calculate the expected
trigger rates as well as the charge and time distributions.

In addition to the ISU work, we have performed complete simulations
using the {\small CORSIKA} shower generator \cite{corsika}, and found
reasonable agreement between the two Monte Carlos. All simulation
results shown in this paper are from ISU (with the exception of
figure \ref{tdist}).
 
\section{STUDY OF AIR SHOWERS WITH A SIX-HELIOSTAT ARRAY}
We took data during the December, January, and
February 1996-1997 new moon periods. The results presented in this
article use the February data, recorded with the Flash ADC's.
The trigger multiplicity was 5 heliostats out of 6 above threshold and the single PM threshold was 5 photoelectrons.
To estimate the charge in an event (figure \ref{typ_evt}) we sum 4 samples before the peak and
11 samples after the peak.
The pulse arrival time on a single
channel is taken as the voltage-weighted mean over the same samples.
The selection cuts we use in the analysis are a requirement of a minimum peak height of four  Flash
        ADC counts for five out of the six channels corresponding to 7 photoelectrons, without pedestal
        subtraction.

\begin{figure}
\centerline{\includegraphics[scale=0.8]
{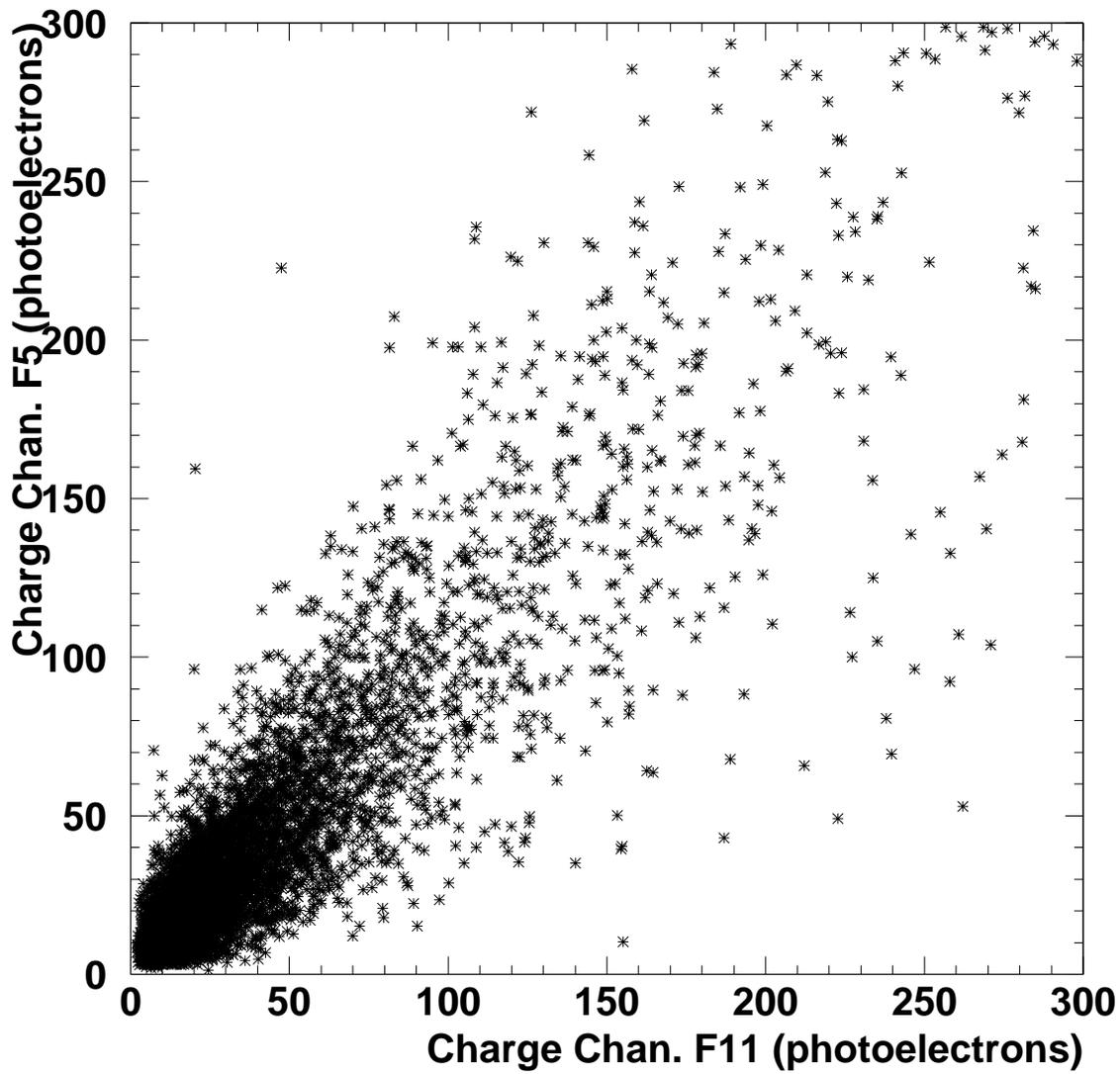}}
\caption{  Phototube pulseheight in heliostat F5 versus the pulseheight
in heliostat F11, for six-fold trigger coincidences while pointing at
10 km, recorded using the Flash ADC's.}
\label{chg_corr}
\end{figure}

\begin{figure}
\includegraphics[scale=0.8]{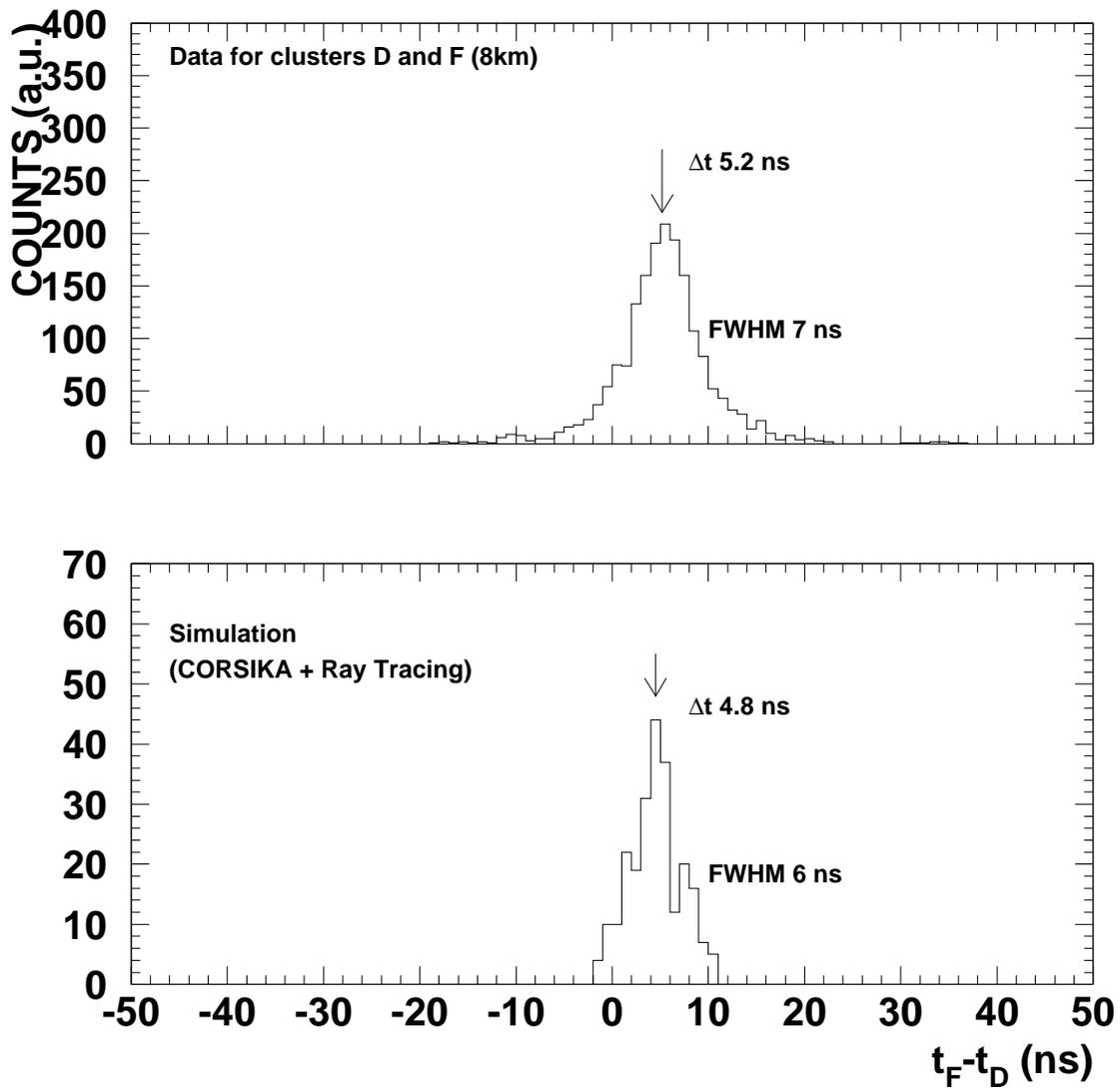}
\caption{  Difference of the average arrival time of 
the Cherenkov pulse on the F and D heliostats, when pointing
at 8 km above heliostat D2. Top: data. Bottom: Corsika Monte Carlo
simulation.}
\label{tdist}
\end{figure}

A variety of evidence convinces us that our data is dominated
by Cherenkov light pulses from extended air showers.
For example, figure \ref{chg_corr} shows a strong correlation
between the pulseheights observed in heliostats F5 and F11.
Accidental coincidences between random phototube noise
({\em e.g.} night sky or afterpulses) would be uncorrelated.
Similarly, the timing correlation within the 20 ns coincidence
gate (shown in figure \ref{alt_time}), the ``break'' 
in the coincidence rate curve near 5 photoelectrons (figure \ref{sum_rate})
and the good agreement between the observed and predicted differential
charge distributions (figure \ref{diff_chg}) provide further
evidence. We will discuss these arguments in the following sections. 

{\em Convergent pointing: }
Equation \ref{thresh_form} underscores the necessity of a small
field-of-view in order to achieve a low energy threshold. A consequence
of a small {\em fov} is that widely-spaced heliostats aimed at a
gamma-ray source will in general {\em not} all see Cherenkov photons
from a given atmospheric cascade. Stated differently, to optimize
photon collection efficiency the heliostats
must track the zone in the atmosphere where the Cherenkov light
is generated. In the tests
all heliostats were pointed at a point $z$ kilometers above
heliostat D2 (see figure \ref{principle}). We varied $z$ from 3 to 100
km.

\subsection{Timing measurements}
Figure \ref{tdist} shows the distribution of the differences
of the arrival times averaged over the F and the D heliostats.
The top plot shows measurements made with the Flash ADC's, while
the bottom plot was made using the {\small CORSIKA} Monte Carlo.
The width is due mainly to the thickness of the Cherenkov wavefront
for hadron showers: the dispersion due to the zone defined by the
overlap of the fields of view of the two groups of heliostats is
of the order of 3 nanoseconds. The 0.4 ns difference between the measured and
calculated mean times are due in part to accumulated uncertainties
in the cable delay measurements, as well as to a small uncertainty
in the heliostat pointing arising from the method we used to place the
phototube in the secondary mirror focal plane.

\begin{figure}
\includegraphics[scale=0.9]{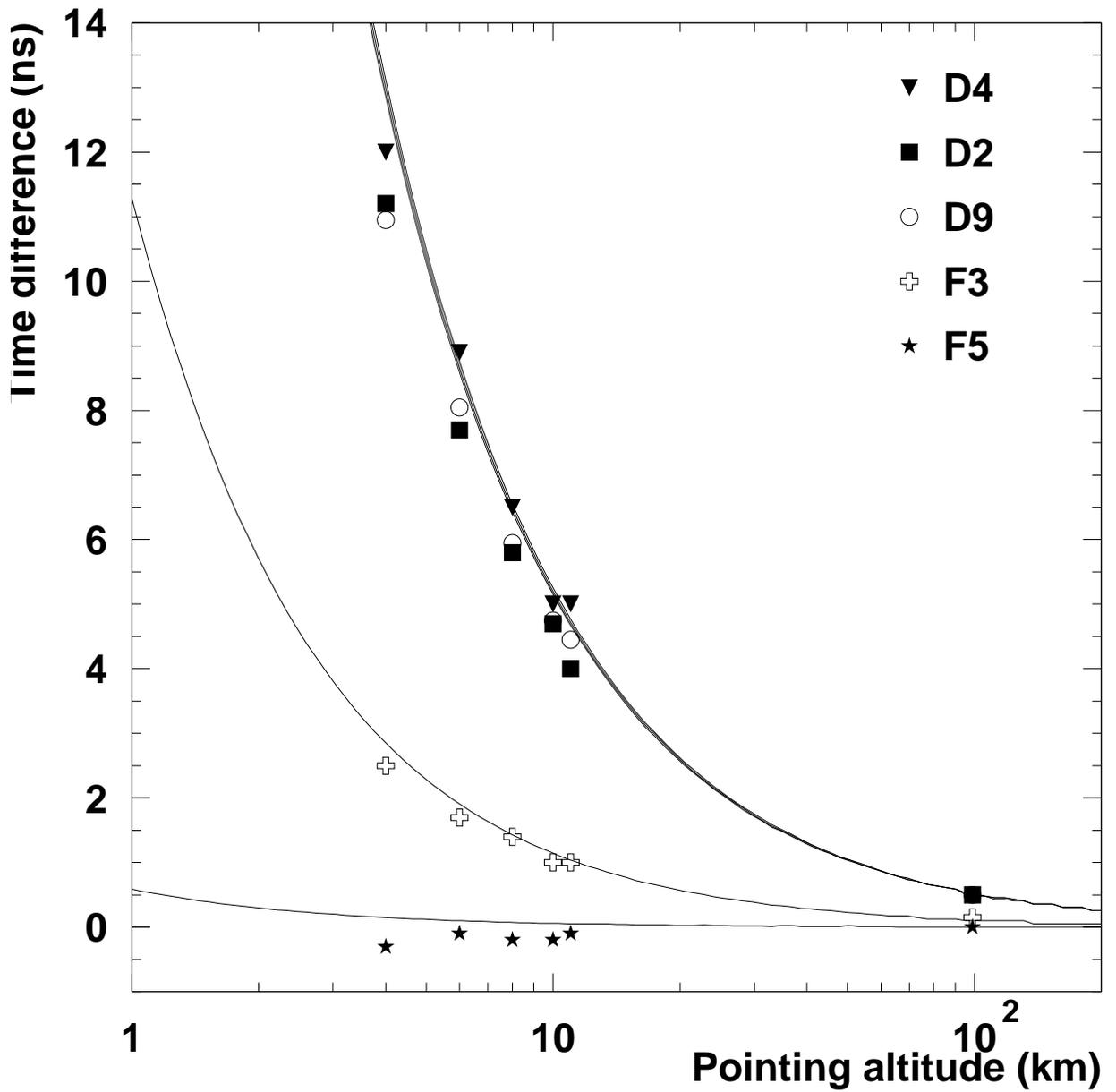}
\caption{  Cherenkov pulse arrival times with respect to F11, 
as a function of
pointing altitude (above the site), for different heliostats.}
\label{alt_time}
\end{figure}

\begin{figure}
\includegraphics[scale=0.8]{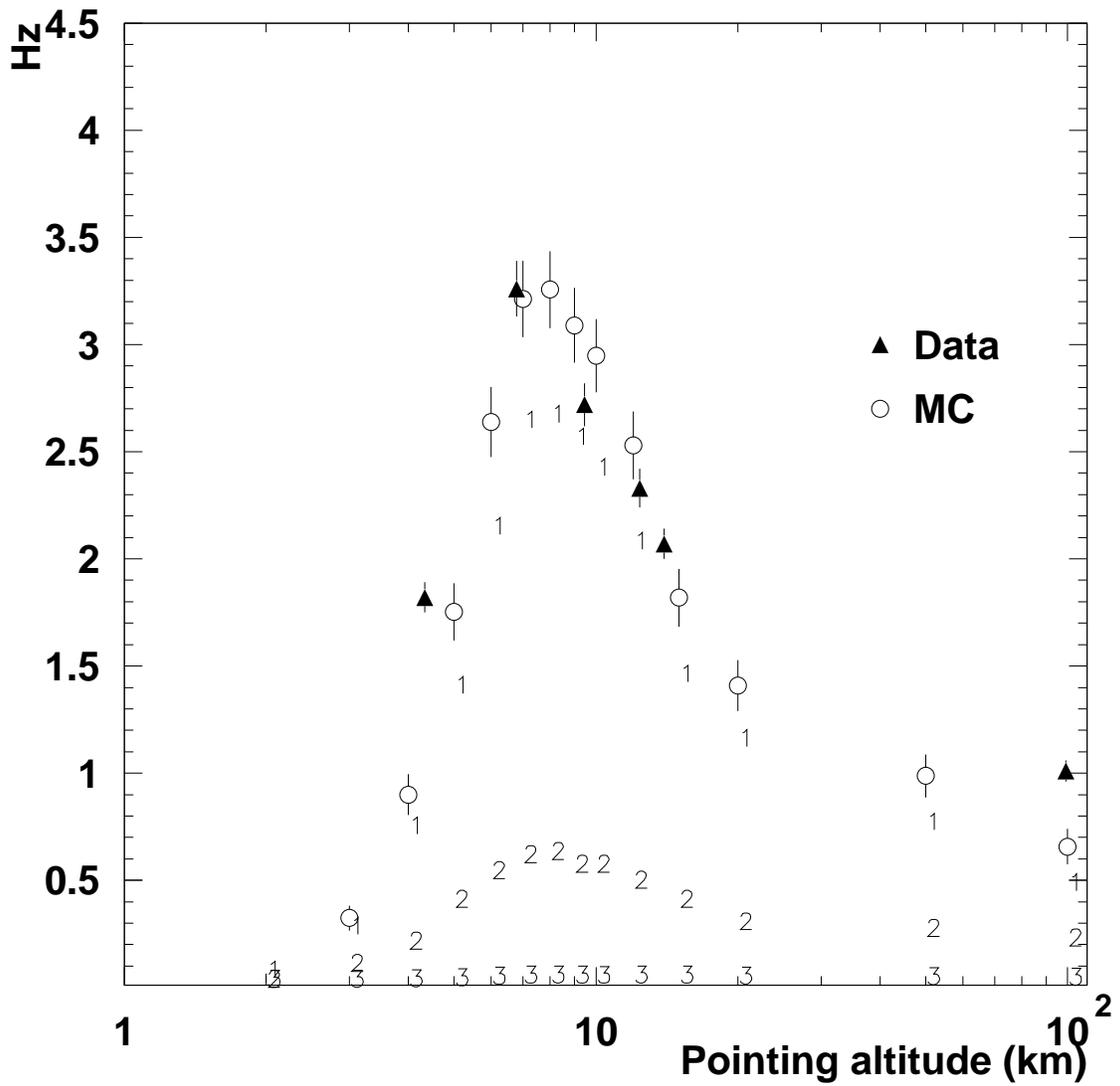}
\caption{  Counting rate versus pointing altitude above the
detector after data
analysis cuts. Triangles: Data. Circles: sum of the Monte Carlo
H (1), He (2), and Li (3) contributions.}
\label{altitude}
\end{figure}

Figure \ref{alt_time} summarizes the measurements
of the arrival time difference for Cherenkov light as a function of 
pointing altitude.
The points are the means of the distributions
of the differences between the arrival time for each heliostat and
the arrival time for F11.

The curves plotted on the figure come from a simple model, where
we assume that the light is emitted from the point that the heliostats
aim at, and the time difference is then
\begin{equation}
c\Delta T_i = \sqrt{z^2+d^2_{i}} - z,
\end{equation}
where $d_{i}$ is the  distance  between  heliostat  $i$  and  F11.  This
        particular heliostat, the farthest from the observed sector in the sky, was chosen to give positive time differences
        in figure \ref{alt_time}.
We imposed agreement between the simple model and the data
at 99 km, to correct for the 0.4 ns offset described
in the preceding paragraph.
The good agreement between the data and the simple model
demonstrates the ability to point the telescope accurately and
to measure nanosecond time differences accurately, which are
two critical aspects of the performance of a Cherenkov gamma-ray telescope.

\subsection{Trigger rates}
\label{trg_sect}

{\em Rate versus pointing altitude: }
 The pointing altitude
that optimizes Cherenkov light collection for 20 GeV gamma-ray showers
is approximately 13 km above the Themis site.
Hadron showers are harder to model, and a measurement
of the hadron shower trigger rate as a function of pointing
altitude is useful to choose the pointing strategy that will
optimize the gamma/hadron signal-to-noise ratio. 
Figure \ref{altitude} summarizes our measurements. In this particular study the multiplicity was 6 heliostats out of 6 above threshold.
The numbers correspond to the charge of the H, He, and Li nuclei
that we simulated. The sum of the three is shown by the circles.
The absolute rate observed in the data (triangles) 
agrees well with the Monte Carlo prediction.

{\em Rates versus trigger threshold:}

\begin{figure}
\includegraphics[scale=0.8]{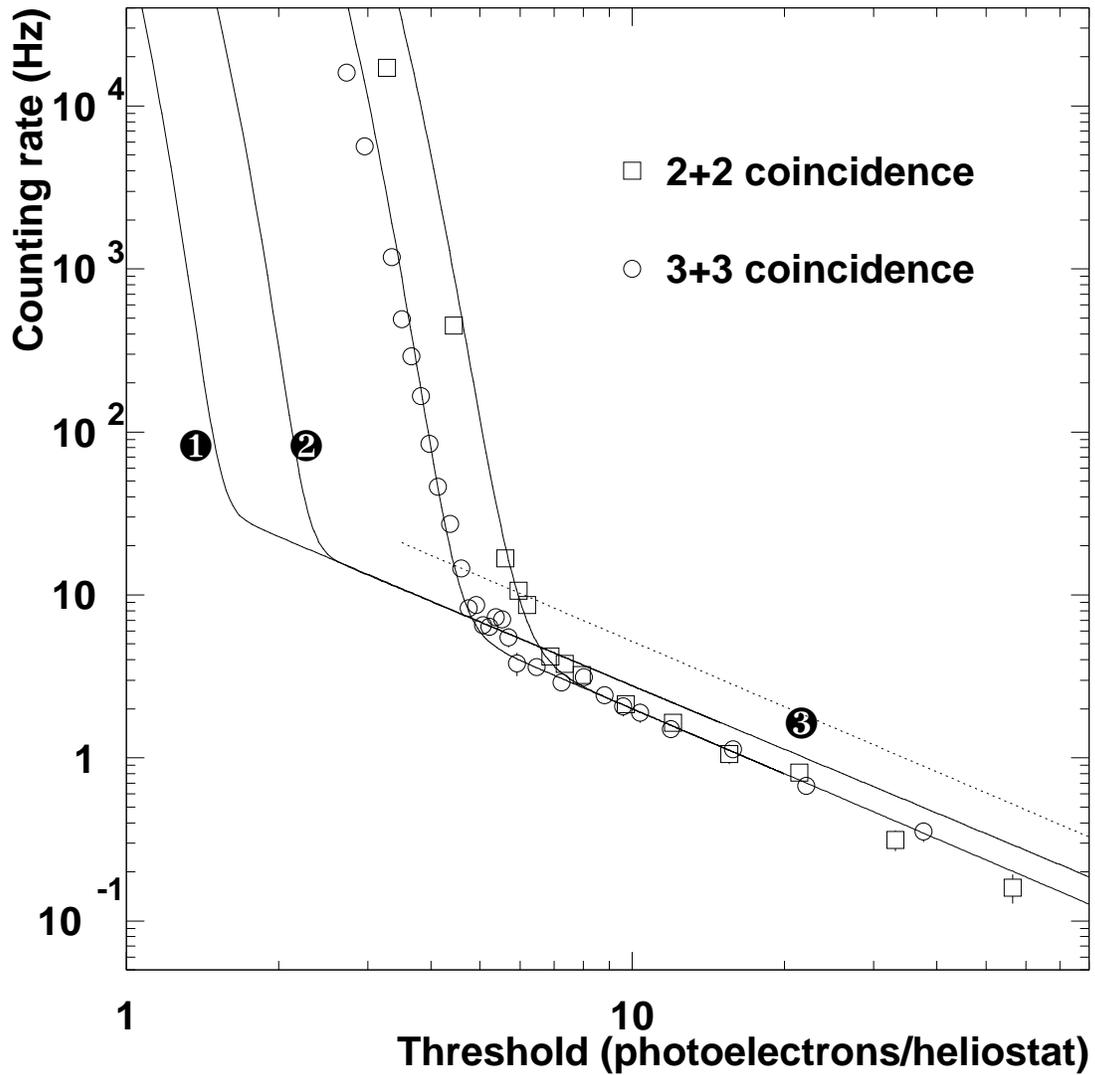}
\caption{  Counting rate versus trigger threshold using a coincidence
of two analog sums. Circles: Coincidence of two sums of three phototubes.Squares: Coincidence of two sums of two phototubes. Curve number 1 (2): expected rates for respectively 3 (4) groups of 6 (9) heliostats. See the text for the improved level of dotted line 3.}
\label{sum_rate}
\end{figure}

Figure \ref{sum_rate} shows the counting rate as a function
of threshold 
as measured using a prototype of the hybrid analog/digital
trigger built for the 40-heliostat experiment. The pointing altitude  is
        10 km above the site.
A discriminator threshold is applied to the analog sum
of phototube outputs within a subgroup, and the coincidence
rate for two such sums is shown with the circles (corresponding to a sum of three outputs) and  squares (sum of two outputs),
using a 20 ns gate.
The threshold in figure \ref{sum_rate} is normalized to the number of phototubes in the sum (i.e. the threshold is divided by the number of channels combined in the discriminated analogic sum).
The gain used to convert from millivolts to photoelectrons
is the average of the gains of the channels in the sum.
The solid line through the data is the sum of a power  law   fitted to the Cherenkov light from air showers,  and  a
        simulation of the night sky noise.

\begin{figure}
\includegraphics[scale=0.8]{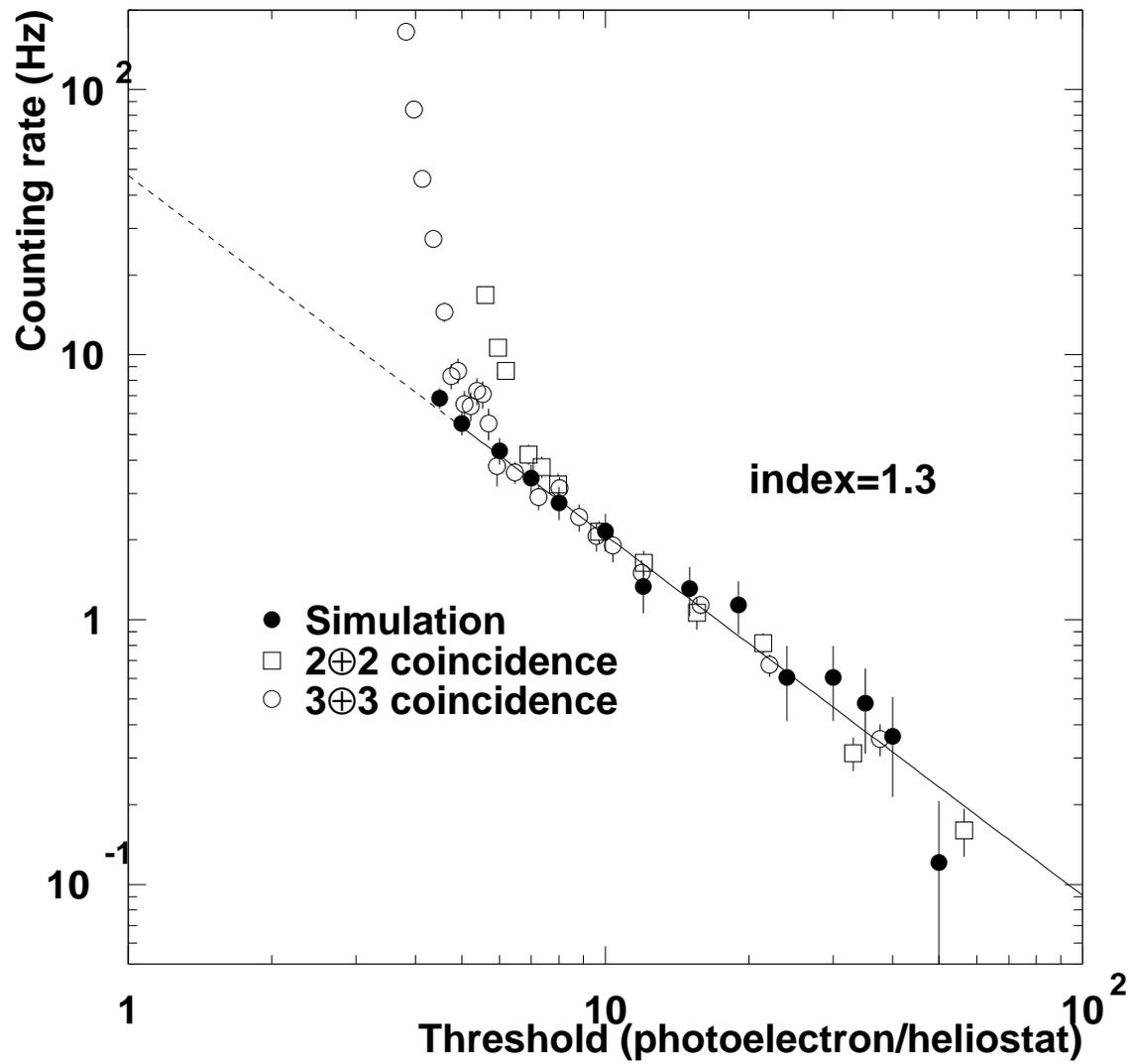}
\caption{  Comparison of Monte Carlo and measured counting 
rates versus trigger threshold for the coincidence
of two analog sums. Open: Data. Solid: Air shower simulation.}
\label{MCsum_rate}
\end{figure}

 Both contributions are well described in terms of known quantities. The night sky light dominates below the 5 photoelectron threshold, and the solid line in this part is the trigger simulation for a photon flux of $b=0.6\gamma _e$/ns (which is slightly above the measurement of section \ref{nsl}). Phototube afterpulsing is taken into account in the trigger simulation.We tested that coincidences at thresholds below the
break point are due mainly to accidental overlap of random
night sky photons by increasing the 20 ns discriminator output
to 40 ns, and we verified that the observed rate at fixed threshold
doubled, as expected for accidentals. Changing the coincidence of a sum of two heliostats to a sum of three heliostats affects significantly the night sky light induced rate, as shown by the open markers in figures \ref{sum_rate} and \ref{MCsum_rate}, but not  the air shower part.

Above the 5 photoelectron breakpoint, the $3 \oplus 3$ and $2 \oplus 2$
data lie on a curve of the form $f_C(s)=40s^{-1.3}$ predominately due to Cherenkov flashes, where $s$ is the threshold in photoelectrons/heliostat.
The power law index of the integral rate curve is less
than the index of 1.78 inferred from equation \ref{proton_eq}
which is
reproduced by the Monte Carlo as shown in figure \ref{MCsum_rate}.
and is due mainly to the fact that the Cherenkov yield
of a hadron shower, normalized to the primary particle energy,
is decreasing in our energy range.
The rates in figure \ref{MCsum_rate} are
absolute: no {\em a posteriori} normalization was performed.

{\em Extrapolation to the full-scale experiment:} The agreement between the simulated night  sky
        light counting rate and  the  experimental  data  allows  us  to
        estimate with good confidence the expected counting rate for a
        larger setup such as four groups of  9  heliostats  (solid  line
        number 1 in figure \ref{sum_rate}) or three groups of 6 heliostats (solid line number 2).
        The power law above a two  photoelectron threshold is
        the  result  of  a  detailed  Monte  Carlo  of  the   full-scale
        experiment. The dotted line (number 3) is  an  extrapolation  of
        the measured $3 \oplus 3$ proton trigger rate to the full-scale experiment, using mainly the fact  that
        the field-of-view will be different. The simulated counting rate
        of  the full-scale experiment  is lower  than the extrapolated counting
        rate. This is due to the fact that a third group of
        heliostats in the trigger requires a more  uniform light
        pool, thus improving hadron rejection at the trigger level.

        One of  the  conclusions  of  figure \ref{sum_rate}  is  that  a  trigger
        threshold of about  two  photoelectrons, with an  acceptable  raw
        trigger  rate   is   realistic   for   the full-scale
        experiment. This threshold should  provide  a  gamma-ray  energy
        threshold of about 20 GeV.

{\em Muon rates:}
Muons falling near a group  of  heliostats  are  not  detected  because
        the light is not emitted in the volume defined by the intersection
        of the field-of-view of the two groups.  We
        generated $10^6$ muons above 10 GeV  according  to the measurements of
        \cite{muons}, corresponding roughly to 75 minutes  of  exposure.
        Ten muon events triggered for a simulated pointing altitude  of
        6 km, from which we deduce that the muon rate is less than 4 mHz
        at the 95\% confidence level. The 6 km altitude  is
        probably due  to  the  fact  that  the  angle  between  the  two
        field-of-views of the two groups of heliostats then corresponds to the
        altitude dependent Cherenkov angle.

\subsection{Photoelectron distributions and energy thresholds}

\begin{figure}
\includegraphics[scale=0.8]{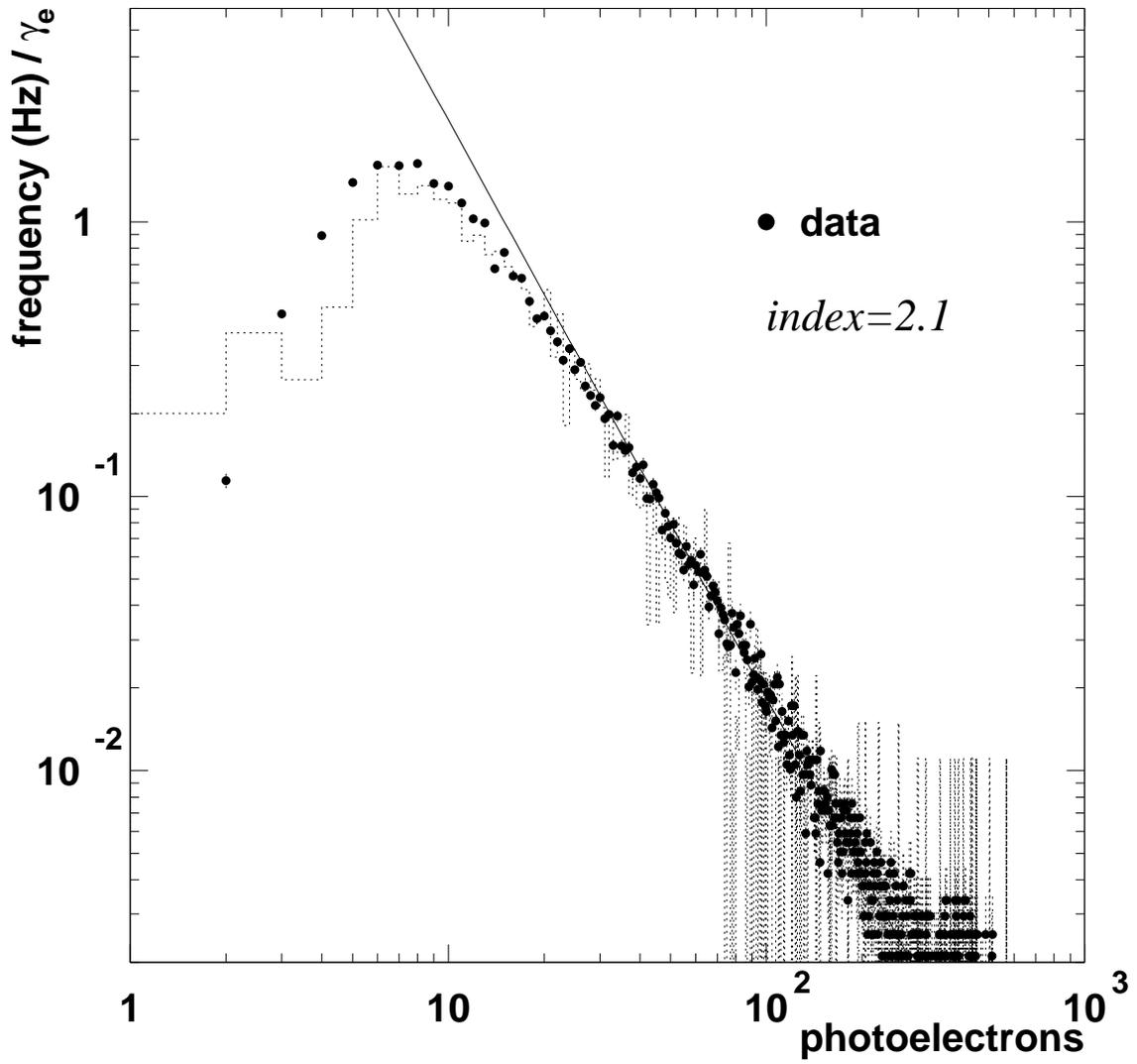}
\caption{  Differential charge distribution of all channels expressed in photoelectrons,
 while pointing at 10 km above the site. The histogram is the Monte Carlo prediction.}
\label{diff_chg}
\end{figure}

Figure \ref{diff_chg} compares the differential photoelectron distribution
with the prediction of the Monte Carlo simulation, for 15000
events recorded at a pointing altitude of 10 km. 6491 events
remain after selection cuts.  The digitized charge of all channels are combined to build the experimental differential photoelectron distribution. The predicted and measured rates are absolute.
As stated before, a charge threshold of 7 photoelectrons on each channel and a requirement of 5 heliostats out of 6 above threshold was
applied in the data analysis as well as in the Monte Carlo.
The 90\% point on the leading edge of the differential
plot is at 6 photoelectrons, in good agreement with the
trigger and analysis thresholds.

Figures \ref{MCsum_rate} and \ref{diff_chg} demonstrate the reliability of our detector simulation:
our modeling of the detector response to atmospheric Cherenkov
light reproduces our observations faithfully. The simulations correctly reproduce various parameters and distributions obtained from the experiment. We can therefore
use our simulation to address further questions with confidence.

\begin{figure}
\includegraphics[scale=0.8]{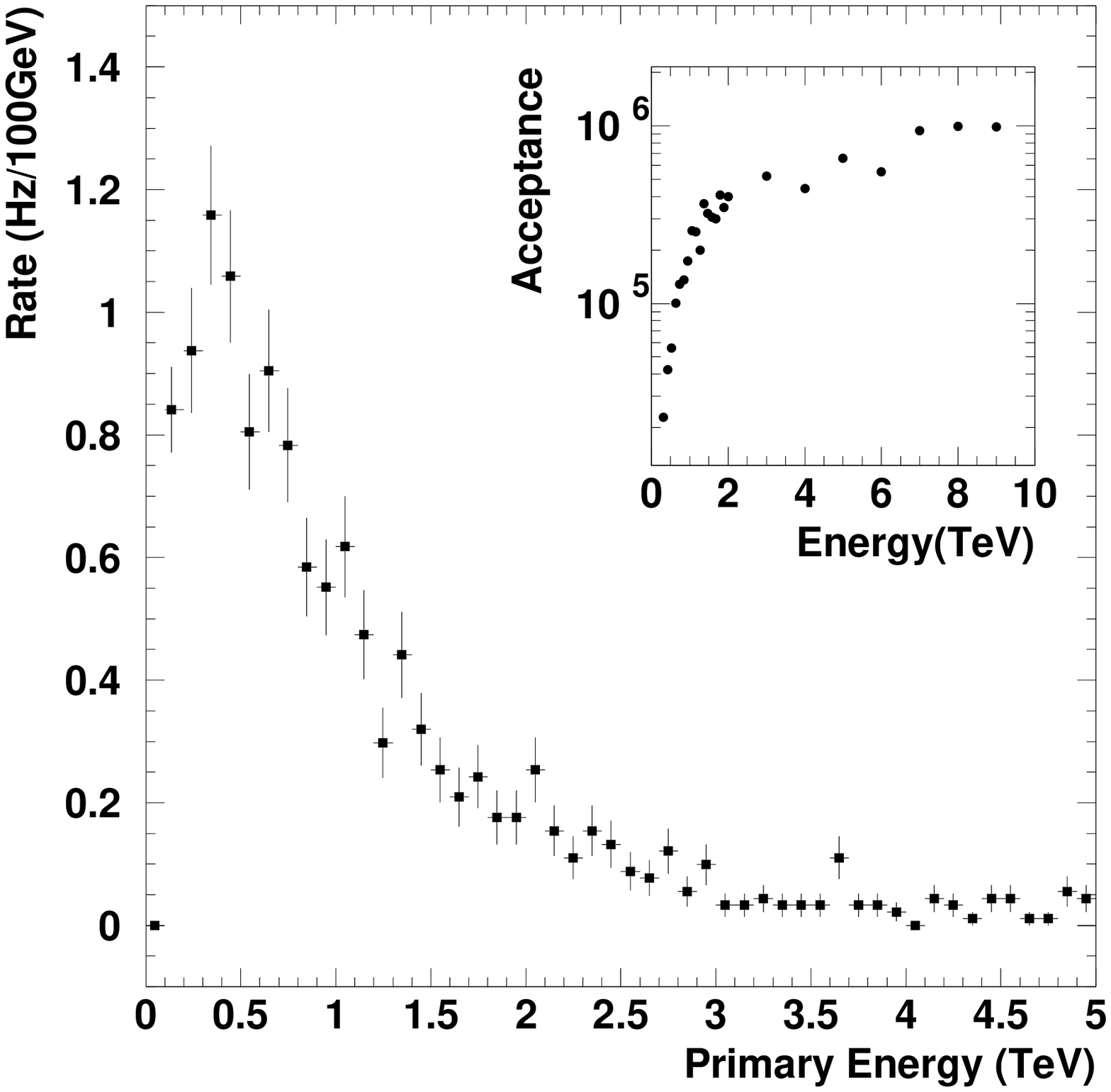}
\caption{  Monte Carlo differential rate of reconstructed showers as
a function of primary proton energy, for a pointing altitude 10 km
above the site and normalized to the input spectrum. The single channel threshold is 7 photoelectrons. 
The energy threshold is near 300 GeV. Inset: the acceptance in units of $cm^{2}sr$.}
\label{seuilen1}
\end{figure}

{\em Hadrons:} 
Figure \ref{seuilen1} shows the Monte Carlo differential
energy distribution for the proton triggers of the simulated histogram in
figure \ref{diff_chg}. The energy threshold is near 300 GeV.
The hybrid trigger provides a comparable but slightly
lower threshold.
The acceptance in the inset to figure \ref{seuilen1} is in units of
$cm^2\textrm{ }sr$.

{\em Gamma-rays:}
Simulation of gamma-ray events in the same experimental conditions
yields a gamma-ray threshold of 80 GeV. In the 40-heliostat experiment
the Cherenkov light collection efficiency will be better, due to 
improved secondary mirrors and Winston cones.
The gamma-ray energy threshold will thus be lower than that suggested
by a naive extrapolation based on the curves in figure \ref{sum_rate}.

\section{CONSEQUENCES FOR FUTURE LOW ENERGY ATMOSPHERIC CHERENKOV TELESCOPES}
Two different approaches to ground-based detection of 20 to 200 GeV gamma-ray detection are being worked on.
The {\small MAGIC} project aims to extend the imaging technique below the
current minimum of about 200 GeV through greatly improved Cherenkov 
light collection \cite{magic}, using a 17 m diameter mirror.  They
        are developing a camera that would have over 50\% quantum efficiency. 
The other approach is the sampling technique of {\small STACEE} 
\cite{stacee} and {\small CELESTE}, which both 
rely on existing solar plants. 
The design of a full-scale instrument sensitive 
in the 20 to 200 GeV range must wait for the results of these projects. 
Even so, the results presented in this paper allow us to underscore
some elements of the choice between a sampler and an imager.

\subsection{Trigger level background rejection}
The Crab detections by Themistocle \cite{them} and 
{\small ASGAT} \cite{asgat} established the validity 
of wavefront sampling with a multi-telescope array.
Sampling is less efficient than the imaging technique in the TeV domain: 
the angular precision is similar, but hadron rejection is worse. 
At lower energies the situation changes. The Cherenkov light 
yield for hadron showers below 100 GeV decreases faster than the 
energy, so that false triggers for imaging devices will be due mostly to muons 
falling at distances up to several mirror radii. 
A sampler triggers on simultaneous signals from 
distant telescopes. This trigger selects showers with a
uniform light pool, such as gamma-ray showers or hadron showers that have fragmented
mainly to a neutral pion. In particular, muons do not trigger the device.

Therefore an important result of the present study is the low trigger rate obtained
at low energy threshold, as for example in figure 11. 
Extrapolation to the 40-heliostat array predicts about 20 Hz for a fraction
of a Hz of gamma events from the Crab.  
The large imager project {\small MAGIC} expects a raw trigger rate of order 1 kHz for the same
energy range. This high rate may be reduced by online and offline data reduction \cite{magic}.  
This will be better understood when data are available. At present, we can only speculate about how a sampling array might look.
               
\subsection{Towards a major low energy atmospheric Cherenkov detector}
The solar arrays permit low-cost full-scale testing of the sampling
approach at low energy. 
Phototubes on each mirror as for Themistocle or 
{\small ASGAT} is better. 
The disadvantages of the solar tower set-up are:\begin{itemize}
\item{ the incident angle of the Cherenkov light on the heliostats varies widely while tracking a cosmic source. A large incidence angle
degrades  the  effective  collection  area  and  the  light   collection
        efficiency decreases with the increasing aberration effects. 
Conventional telescopes look straight at the shower while tracking 
the cosmic source and always work near their optical axis;} 
\item{ the field-of-view must be less than 1 degree because of the limited 
space at the top of the tower. 
The small {\em fov} is exploited to increase the signal-to-noise (Cherenkov
to night sky light) ratio, by pointing the heliostats at the 
shower core (see section 3). This strategy
limits the collection area to about 2$\times$10$^4$ m$^2$ depending on the field-of-view. 
This shortcoming can be solved with conventional telescopes by using a 
camera with several pixels:  a central phototube can provide trigger
information, while a ring of phototubes would
measure the centering. A similar arrangement is not possible at
the top of a solar tower because of the lack of space.}
\end{itemize}

In short, a sampling array of about the same size as {\small CELESTE},
for example, fifty telescopes of seven-meter diameter, each
could exploit the sampling scheme without the drawbacks of a 
fixed central focus point. Energy threshold and overall
performance could be directly extrapolated from {\small CELESTE}'s.
This extension of the 
solar plant test bench to an optimized sampling array  - on  an  optimal
        site - might be the best towards a major low energy  atmospheric
        Cherenkov detector.  A clear consequence of the present 
study is that such a scheme is at least feasible.

\section{CONCLUSIONS}

We have used a solar array to detect Cherenkov light from air showers. We accurately
measure the shower light emitted in the small volume at the intersection
of the fields-of-view of two groups of three heliostats each. 

Triggering on widely spaced heliostat groups has been shown to
give good hadron and muon rejection.

The minimum trigger threshold at which the Cherenkov signal dominated
the night sky background was 5 photoelectrons per heliostat. 
In our analysis the threshold was 7 photoelectrons, which
corresponds to a hadron shower energy slightly above 300 GeV, 
or to a gamma-ray energy of 80 GeV. At this threshold the 
trigger rate was 6 Hz. We expect a threshold
of 2 -- 3 photoelectrons and a rate of 20 Hz with the optics now being installed, 
corresponding to a gamma-ray energy threshold below
30 GeV.
 
Our test results, and in particular the low trigger rate,
confirm that wavefront
sampling is an alternative to imaging as
a basis for a high-performance gamma-ray detector in the 
20 to 200 GeV energy range. We are presently commissioning 
a full-scale device which we expect to provide the first ground-based
detection of cosmic gamma-rays in this energy range. The ongoing observations 
could establish that, even independently of existing solar arrays, wavefront sampling
        might be the best way to access 
an  unexplored  spectral  window  known  to  be  particularly  rich   in
        astrophysical information.
\\

 {\bf ACKNOWLEDGEMENTS}\\
We gratefully acknowledge the contributions of the technical staffs
of our laboratories and the IMP (CNRS) at Odeillo.
Drs. Philippe Roy and Louis Behr play key roles in 
the continued progress of the project.
We thank Electricit\a'e de France for allowing
use of the Themis site. Funding was provided by the Institut National
de Physique des Particules et de Physique Nucl\a'eaire of the Centre
National de Recherche Scientifique; the Grant Agency of the Czech
Republic; by the University of Bordeaux; by
the Ecole Polytechnique; and by the Regional Councils of 
Aquitaine and of Languedoc-Roussillon.

\newpage

\clearpage

\clearpage

\clearpage

\clearpage

\clearpage

\clearpage

\clearpage

\clearpage

\clearpage

\clearpage

\clearpage

\clearpage

\clearpage

\begin{thebibliography}{9}

\bibitem{proposal} {\small CELESTE} experimental proposal. 
See
http://wwwcenbg.in2p3.fr/Astroparticule or contact the author.

\bibitem {vonM}C. von Montigny {\it et al}, Ap. J. {\bf 440} (1995) 525.

\bibitem{Egpuls}D.J. Thompson {\it et al}, Ap. J. {\bf 465} (1996) 385.

\bibitem{ir} 
O.C. De Jager, F.W. Stecker, \& M.H. Salamon, Nature {\bf 369} (1994) 294.

\bibitem{weekes} T.C. Weekes, Phys. Rep. {\bf 160} (1988).

\bibitem{whipple}J.  Quinn {\it et al}, Ap. J. Lett. {\bf 456} (1996) 83.

\bibitem{them} P. Baillon {\it et al}, Astropart. Phy. {\bf 1} (1993) 341.

\bibitem{asgat} P. Goret {\it et al}, A\&A {\bf 270} (1993) 401.

\bibitem{cat}
R.J. Protheroe {\it et al} in {it Highlights of the XXV$^{th}$ Int. Cosmic Ray
Conf.}, Durban (1997).

\bibitem{Egcat}D.J. Thompson {\it et al}, Ap.J.S. {\bf 101} (1995) 259.


\bibitem{mini}
Mini-circuits PSC-3-1W, P.O. Box 350166, New York 11235-0003.


\bibitem{stacee}
M.C. Chantell {\it et al}, EFI preprint 38-97 (accepted by NIM).\\
R. Ong {\it et al}, Astropart. Phys. {\bf 5} (1996) 353.

\bibitem{rasmik}
R. Mirzoyan and E. Lorenz, 
{\it proc. ``Towards a Large Atmospheric \v Cerenkov Detector-IV''}, Padova 1995.

\bibitem{isu}
M. Kertzmann and G. Sembronski, Nucl. Instr. Meth. {\bf 343A} (1994) pp.629-643.

\bibitem{corsika}
J.N. Capdevielle {\em et al} Karlsruhe KfK report 4998 (1992).

\bibitem{masakatsu}
Masakatsu Ichimura {\em et al}, Phys. Rev. {\bf 48D} (1993) 1949.

\bibitem{barbara}
B. Wiebel-Sooth, P. Biermann, and H. Mayer, {\it accepted by }
A\&A. Preprint astro-ph/9709253.

\bibitem{magic}
R. Mirzoyan,  E. Lorenz, and Gonz\'ales,
{\it proc. ``Towards a Large Atmospheric \v Cerenkov Detector-IV''}, Durban 1997.

\bibitem{muons}
R. Bellotti {\em et al}, Phys. Rev. {\bf 53D} (1996) 35.



\end{thebibliography}
\end{document}